\documentclass[twocolumn,superscriptaddress,showpacs,pra]{revtex4}
\usepackage{epsfig}
\pdfoutput=1 
\usepackage{epstopdf}
\usepackage{delarray}
\usepackage{amsmath, amssymb}
\usepackage{bm}
\usepackage{graphicx}
\usepackage{placeins}
\usepackage{color}

\begin{document}
\title{Early Detection of the Draupner Wave Using Deep Learning}

\author{Cihan Bay\i nd\i r}
\email{cihanbayindir@gmail.com}
\affiliation{Associate Professor, Engineering Faculty,  \.{I}stanbul Technical University, Maslak, 34467 \.{I}stanbul, Turkey \\
						 Adjunct Professor, Engineering Faculty, Bogazici University, Bebek, 34342 \.{I}stanbul, Turkey \\
						 International Collaboration Board Member, CERN, CH-1211 Geneva 23, Switzerland}

\begin{abstract}
In this paper we propose and apply a deep learning strategy for the early detection of the Draupner rogue (freak) wave, which is also known as the New Year's wave. We use a long short term memory (LSTM) network and show that Draupner rogue wave could have been observed at least minutes before the catastrophically dangerous peak has appeared in the chaotic wave field using the available data. Compared to the existing early warning times scales on the order of seconds, this is a major step forward which will certainly enhance the safety and understanding of the marine engineering. As the rogue wave data sets get improved in future, our results may be enhanced to increase the early warning time scales. Our results can be used to predict other rogue waves and extreme time series phenomena in fields including but are not limited to hydrodynamics and marine engineering, optics, finance and Bose-Einstein condensation, just to name a few.

\end{abstract}
\maketitle

Rogue waves are classified waves having a height more than 2–2.2 times the significant wave height ($H_s$), in the wave field \cite{Kharif}. They are observed in hydrodynamics, optics, quantum mechanics, Bose-Einstein condensation, finance\cite{Kharif, Akhmediev2009b, Akhmediev2009a, Akhmediev2011, BayPLA, BayPRE1, BayPRE2, Wang, Bay_Zeno, Bay_arxNoisyTunKEE, Birkholz}. They are mainly studied analytically and numerically using spectral algoirthms in the frame of some dynamical equations such as the nonlinear Schrödinger equation \cite{Kharif, Akhmediev2011, BayPLA, BayPRE1, BayPRE2, Bay_CSRM, bay2009, Canuto, Demiray2015, Karjadi2010, Karjadi2012, trefethen, Soto2014RwSSchaotic, Baysci, Baytrbz1, Baytrbz2, Peregrine, Bay_arxChaotCurNLS, BayTWMS2015, BayTWMS2016, Bay_TWMS2017, Bay_arxCSRM, Zakharov1968, Bay_cssfmarx}. Although rogue waves may be needed in fiber optics in order to satisfy certain energy levels and locating the information using matched filtering, they are catastrophic in the marine environment and they present a danger to safety of marine operations and their result can be costly. As examples, damaging of the Draupner platform and sinking of the cargo vessel El Faro can be given. In order to prevent such catastrophes, the early detection of rogue waves with precise emergence times and  wave heights is needed in the chaotic ocean. 

There are a couple of existing strategies for the early detection of such waves. Spectral techniques may be utilized by measuring the super-continuum patterns in the Fourier spectra before the rogue wave becomes evident in time \cite{Akhmediev2011, BayPRE2, Bay_arxEarlyDetectCS, Bay_tomog}. However checking the Fourier spectra solely would not give any clue about the expected emergence point (or time) of a rogue wave in a chaotic wave field, thus Fourier spectral analysis is extended to wavelet spectral analysis to locate the emergence point (or time) of a developing rogue wave \cite{BayPLA}. The main weakness of the spectral analysis for the rogue wave prediction is that, the prediction times are on the order of seconds only. Later, Birkholz et. al. have proposed using Grassberger-Procaccia nonlinear time series algorithm for the prediction of rogue waves \cite{Birkholz} and have slightly developed this time scale. Although these short time scales may be beneficial for saving lives in the marine environment, the exposure to a rogue wave may not be avoided for such a short time scale.  Recently, Närhi et. al. have proposed the usage of the machine learning techniques for the prediction of the extreme events in fiber optics which looks more promising and may allow for early detection of optical rogues waves minutes before their emergence in the optical fiber \cite{Narhi}.  

In this paper we propose and apply the deep learning for the early detection of the rogue waves and extreme events. With this motivation we use the Long Short Term Memory (LSTM) network, which is one of the very successful deep learning networks, and apply LSTM network to Draupner rogue wave time series for the early detection purposes. We show that, even with the scarce oceanic rogue wave data, the existing early warning times on the order of seconds can be significantly improved up to the order of minutes. We discuss our findings and usability and the benefits of our approach in the coming sections.

\section{\label{sec:level2}Methodology}

\subsection{Review of the Long Short Term Memory} 

\begin{figure}[htb!]
\begin{center}
   \includegraphics[width=3.7in]{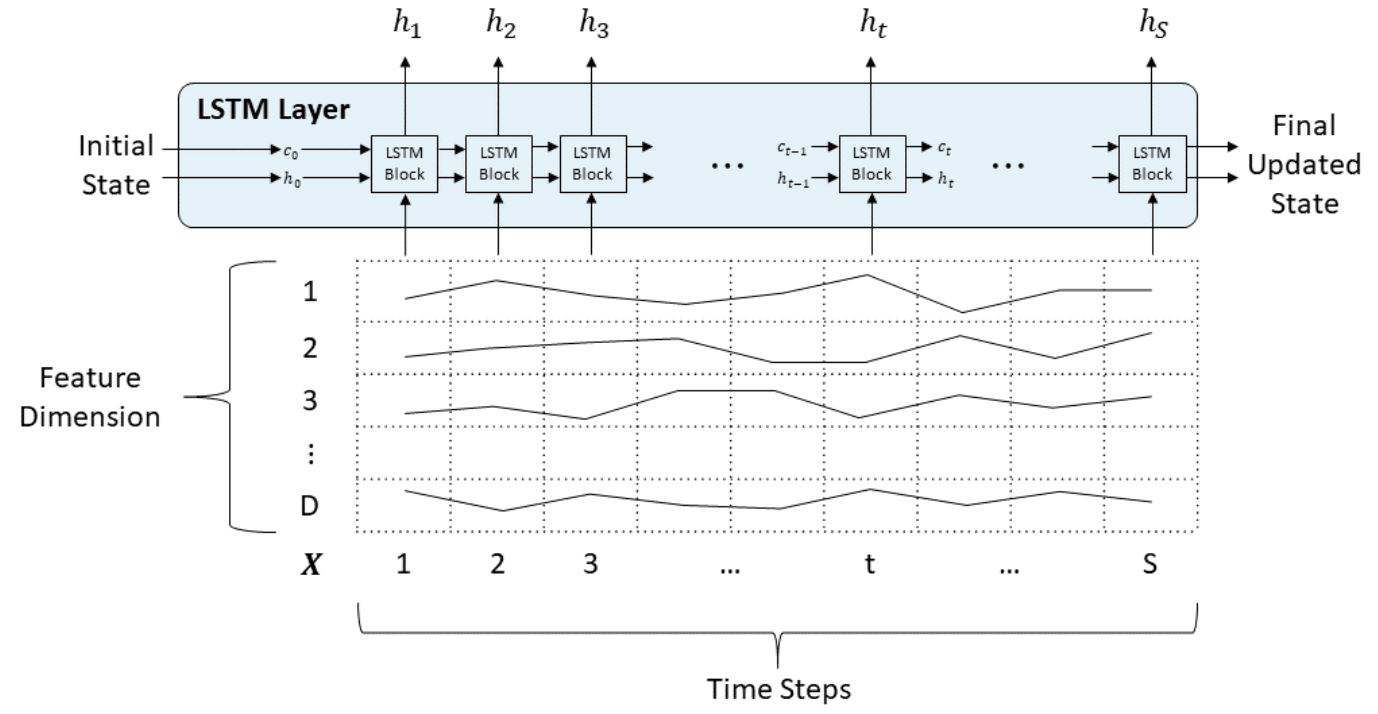}
  \end{center}
\caption{\small LSTM layer architecture \cite{Hochreiter}.}
  \label{figLSTM_layer}
\end{figure}

In this section we give a brief review of the LSTM which was initially introduced in \cite{Hochreiter}. As a deep learning network, one of its uses of the LSTM is its capability in predicting the time series  \cite{Hochreiter, Greff}. To predict the values of future time steps of a given time series,one can train a sequence-to-sequence regression LSTM network. In LSTM network the responses are the training sequences with one time step shifted values \cite{Hochreiter}. Therefore, the LSTM network learns to predict the value of the next time step at each time step of the input sequence \cite{Hochreiter}. An LSTM layer architecture is depicted in Fig.~\ref{figLSTM_layer}.

\begin{figure}[htb!]
\begin{center}
   \includegraphics[width=3.7in]{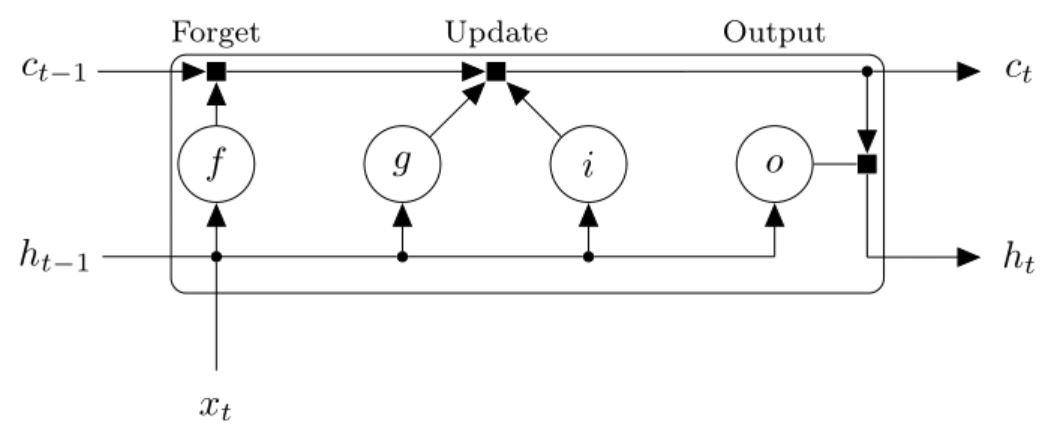}
  \end{center}
\caption{\small LSTM gates \cite{Hochreiter}.}
  \label{figLSTM_gates}
\end{figure}

\begin{figure}[htb!]
\begin{center}
   \includegraphics[width=2.2in]{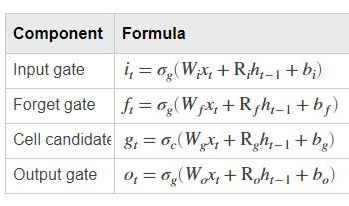}
  \end{center}
\caption{\small LSTM components and formula \cite{Hochreiter}.}
  \label{figLSTM_components}
\end{figure}

The LSTM layer architecture depicted in Fig.~\ref{figLSTM_layer} illustrates the flow of a time series with D features of through an LSTM layer. In here, h denotes the output and c denotes the cell state \cite{Hochreiter}. Starting from the initial time step, the cell state contains information learned at the previous time steps. At each time step, the layer adds or removes information from the cell state. These updates are controlled using the layer consisting of the gates which are depicted in Fig.~\ref{figLSTM_gates} .
In this figure; i, f, g, and o denote the input and forget gates, cell candidate and the output gate, respectively  \cite{Hochreiter}. The cell state at time step t is computed by
\begin{equation}
c_t = f_t \otimes c_{t-1} + i_t \otimes g_{t}
\label{eq01}
\end{equation}
where $\otimes$ denotes element wise multiplication (the Hadamard product)  \cite{Hochreiter}. The hidden state at time t is computed by
\begin{equation}
h_t = o_t \otimes \sigma_c(c_t)
\label{eq02}
\end{equation}
In here $\sigma_c$ is the state activation function for which we use the tanh function. For each time step the functions are summarized in Fig.~\ref{figLSTM_components}
where W indicates the input weight, R indicates the recurrent weights and b indicates the bias \cite{Hochreiter}.  $\sigma_g$ is the gate activation function for which a sigmoid function in the form of $\sigma(x)=(1+e^{-x})^{-1}$ is used \cite{Hochreiter}. Implementing this approach, we first standardize the training data to have zero mean and unit variance for a better fit and we specify the LSTM layer to have 200 hidden units. We set the training epochs option to 250 and in order to prevent the gradients from diverging, we have used a the gradient threshold value of 1. We start the predictions with an initial learning rate 0.005, and the drop it to 0.004 after 125 epochs. We then unstandardize our prediction time series using the parameters mentioned earlier. Additionally we update the network state with observed values instead of predictions by using the time steps between predictions. All LSTM network based predictions in this paper are performed using the parameters summarized above. Our paper is on discussing the usage of the deep learning for the prediction of rogue waves, thus, for a more comprehensive discussion and the details of the LSTM network the reader is referred to \cite{Hochreiter, Greff}.

\section{\label{sec:level3}Results and Discussion}
 
\subsection{Results for the Time-Reversed Draupner Wave Time Series}

In order to check the performance of deep learning on predicting rogue wave phenomena, we apply the LSTM network to the Draupner data in this section. The original Draupner wave record is depicted in Fig.~\ref{fig9} and the details of the time series can be seen in \cite{Haver}. Since the dangerous peak appears within this first 20\% of the original record and we want to analyze the performance of LSTM network for a longer training data range, we first analyze the time-reversed Draupner rogue wave time series. Envelope processes of the water waves are commonly studied in the frame of the nonlinear Schrödinger equation, therefore time reversing the data can mimic the behavior of chaotic wave field at earlier times. Later, we will turn our attention to the original time series as it is recorded.

\begin{figure}[htb!]
\begin{center}
\hspace*{-0.7cm}
   \includegraphics[width=4.1in]{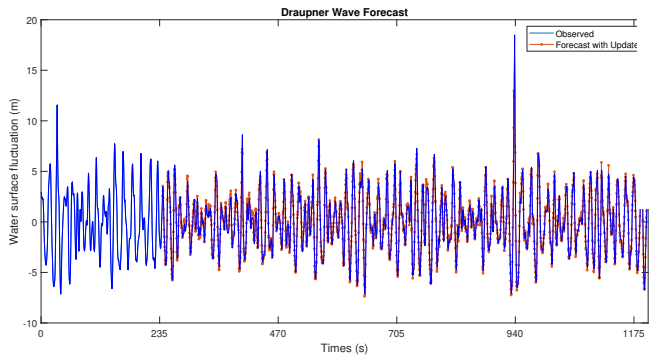}
  \end{center}
\caption{\small Comparison of the full Draupner rogue wave time series and deep learning based prediction using the initial 20\% as the training set.}
  \label{fig1}
\end{figure}

\begin{figure}[htb!]
\begin{center}
\hspace*{-0.7cm}
   \includegraphics[width=4.1in]{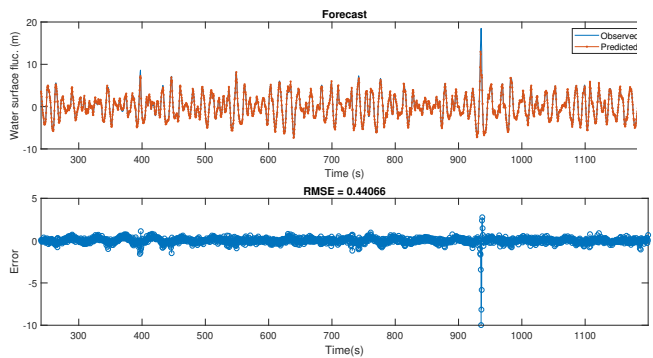}
  \end{center}
\caption{\small Detailed comparison of the observed Draupner rogue wave time series vs time series and rmse error of the predicted time series by the deep learning algorithm that uses the 20\% as the training set.}
  \label{fig2}
\end{figure}

The reversed Draupner rogue wave series is depicted in Fig.~\ref{fig1}, where we use the initial  20\% of the data (depicted in blue) as our training set in LSTM network. The remaining 80\% (depicted in red) is then compared with the observed data in Fig.~\ref{fig2}. As it can be seen in the figure, the LSTM network based deep learning algorithm can predict the emergence of the dangerous peak  approximately 936-240=696s, which is more than 11 minutes,  before it appears. The predicted wave height is around 20.5m, which is slightly less than the recorded peak wave height of approximately 26m. Although the predicted wave height is slightly less than observed, the predicted wave still can be classified as a rogue wave since its height exceeds $2 H_s$.

\begin{figure}[htb!]
\begin{center}
\hspace*{-0.7cm}
   \includegraphics[width=4.1in]{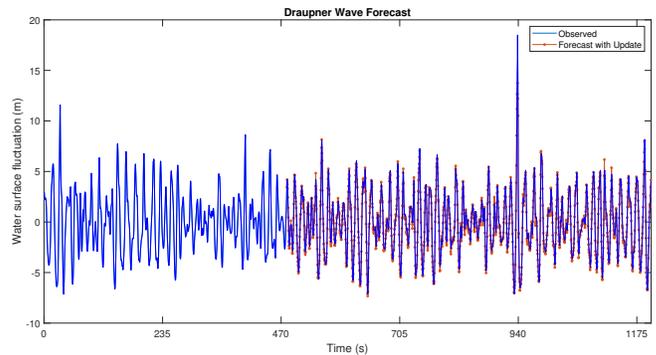}
  \end{center}
\caption{\small Comparison of the full Draupner rogue wave time series and deep learning based prediction using the initial 40\% as the training set.}
  \label{fig3}
\end{figure}

Next, we turn our attention to Fig.~\ref{fig3} and  Fig.~\ref{fig4}, where we depict the results for the case in which initial 40\%  of data is used as the training set of the LSTM network. Coloring of the graphs is as before. Checking the figures one can realize that an early warning time on the order of 936-480=456s, which is more than 7 minutes, is possible. Since the training data is longer, the rms error of the predicted wave height reduces and LSTM deep learning network predicts the dangerous peak wave height as approximately as 21m, which can be classified as a rogue wave as before.

\begin{figure}[htb!]
\begin{center}
\hspace*{-0.7cm}
   \includegraphics[width=4.1in]{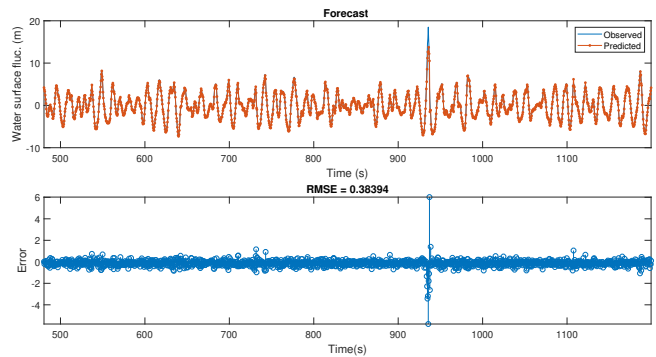}
  \end{center}
\caption{\small Detailed comparison of the observed Draupner rogue wave time series vs time series and rmse error of the predicted time series by the deep learning algorithm that uses the 40\% as the training set.}
  \label{fig4.eps}
\end{figure}

\begin{figure}[htb!]
\begin{center}
\hspace*{-0.7cm}
   \includegraphics[width=4.1in]{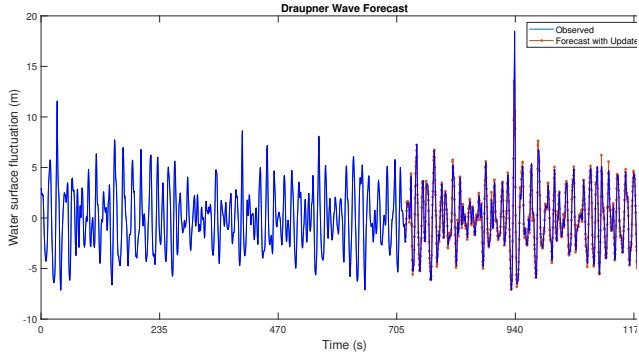}
  \end{center}
\caption{\small Comparison of the full Draupner rogue wave time series and deep learning based prediction using the initial 60\% as the training set.}
  \label{fig5}
\end{figure}

\begin{figure}[htb!]
\begin{center}
\hspace*{-0.7cm}
   \includegraphics[width=4.1in]{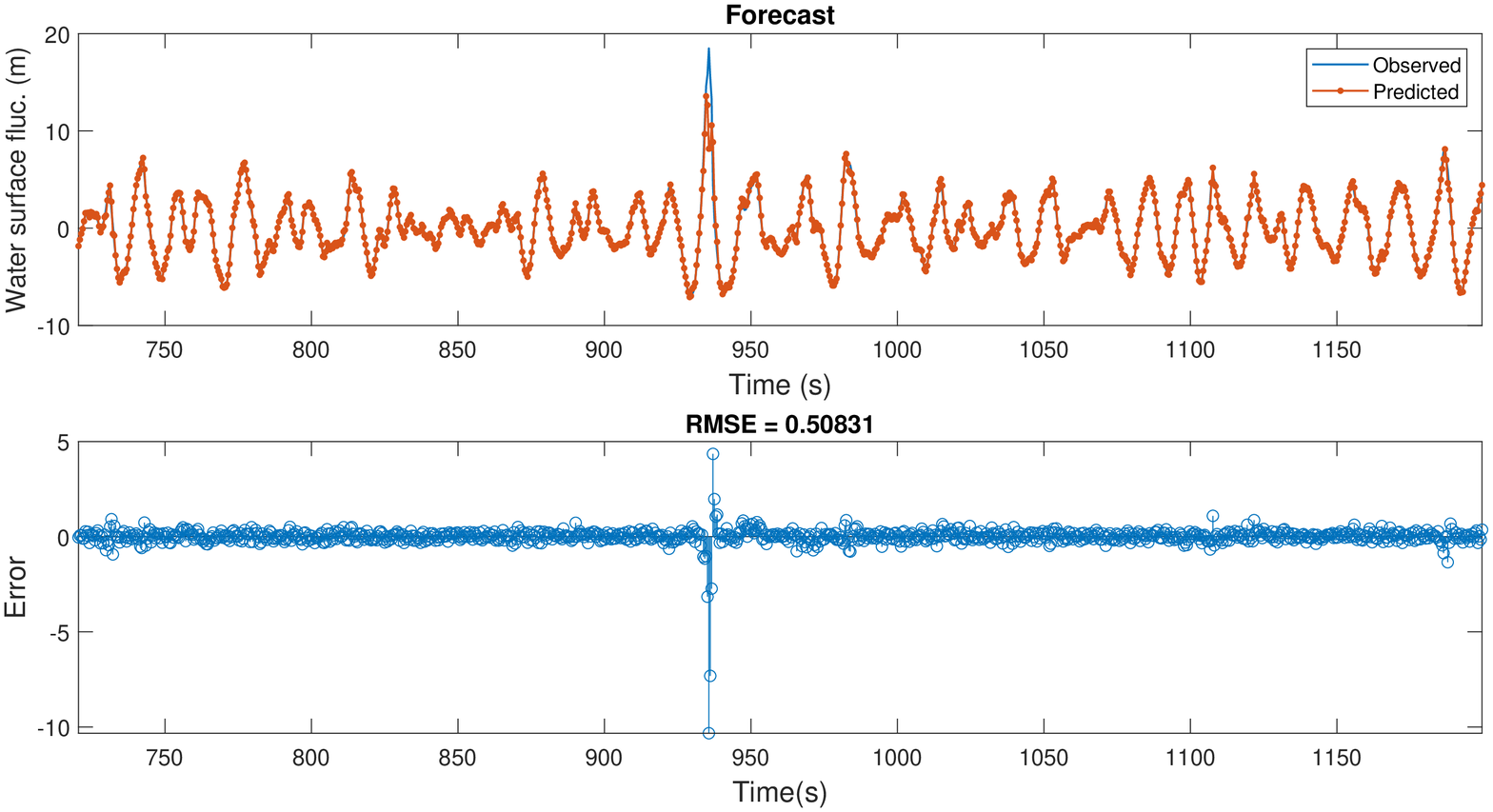}
  \end{center}
\caption{\small  Detailed comparison of the observed Draupner rogue wave time series vs time series and rmse error of the predicted time series by the deep learning algorithm that uses the 60\% as the training set.}
  \label{fig6}
\end{figure}

\begin{figure}[htb!]
\begin{center}
\hspace*{-0.7cm}
   \includegraphics[width=4.1in]{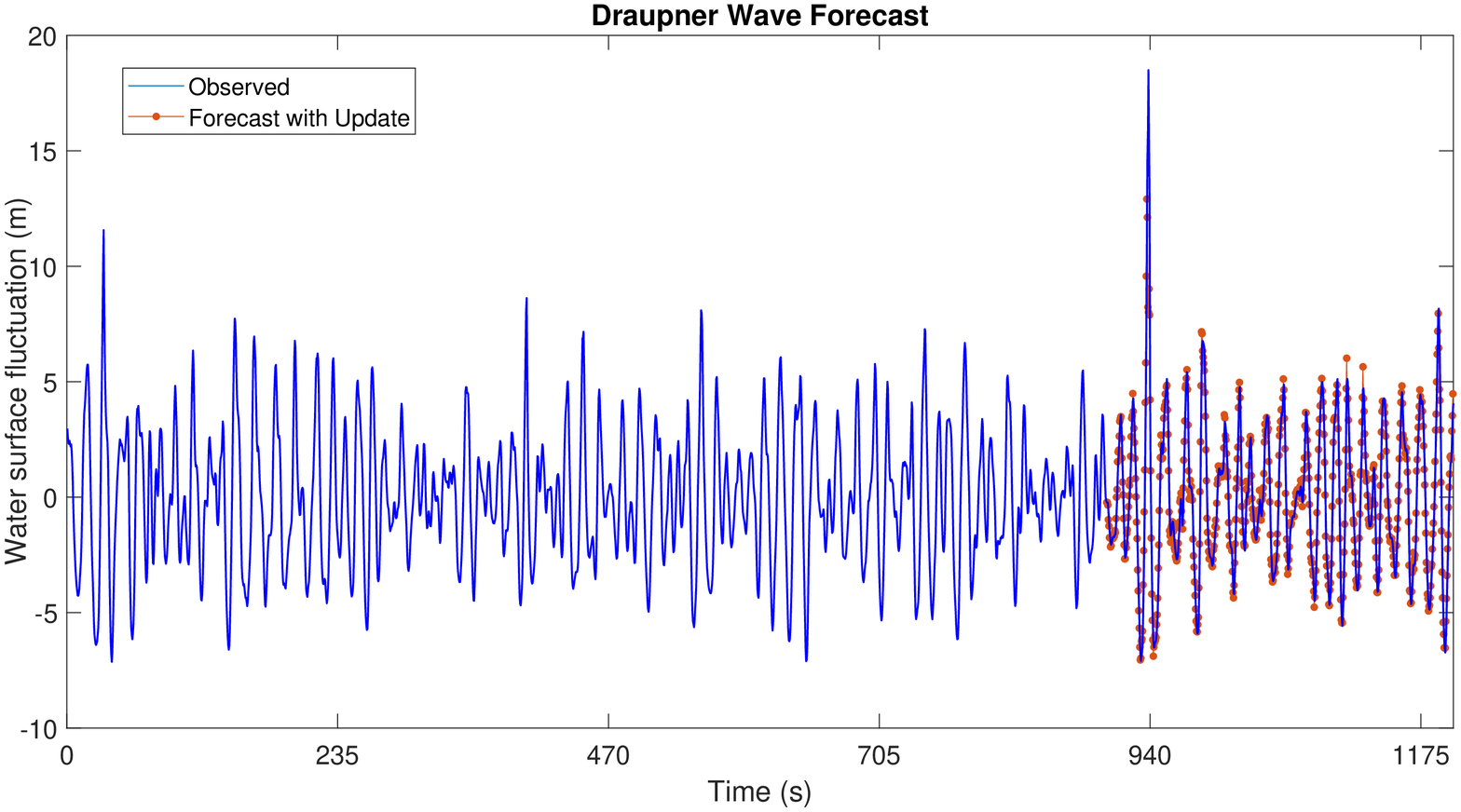}
  \end{center}
\caption{\small Comparison of the full Draupner rogue wave time series and deep learning based prediction using the initial 75\% as the training set.}
  \label{fig7}
\end{figure}

\begin{figure}[htb!]
\begin{center}
\hspace*{-0.7cm}
   \includegraphics[width=4.1in]{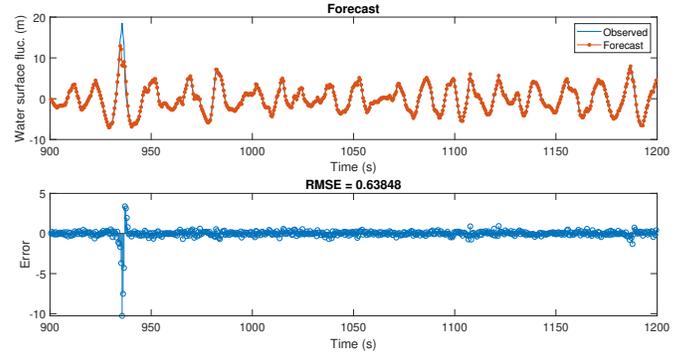}
  \end{center}
\caption{\small  Detailed comparison of the observed Draupner rogue wave time series vs time series and rmse error of the predicted time series by the deep learning algorithm that uses the 75\% as the training set.}
  \label{fig8}
\end{figure}

Similarly, the results for the 60\% training data are depicted in Fig.~\ref{fig5} and Fig.~\ref{fig6}, where as the results for the 75\% training data are depicted in Fig.~\ref{fig7} and Fig.~\ref{fig8}. Checking these figures, one can conclude that an early warning time of 936-720=216s, which is more than 3 minutes is possible for the 60\% training data case and an early warning time of 936-900=36s is possible for the 75\% training case. Although the rms error reduces as the training sequence gets longer, they do not exhibit a linear behavior. Predicted peak wave height is approximately 21m for both of these cases, which is slightly less than the actual observed value of approximately 26m. But again LSTM network based deep learning strategy can be effectively used to predicted unexpectedly anormal extreme waves in the chaotic ocean.

\subsection{Results for the Draupner Wave Time Series}

\begin{figure}[htb!]
\begin{center}
\hspace*{-0.7cm} 
   \includegraphics[width=4.2in]{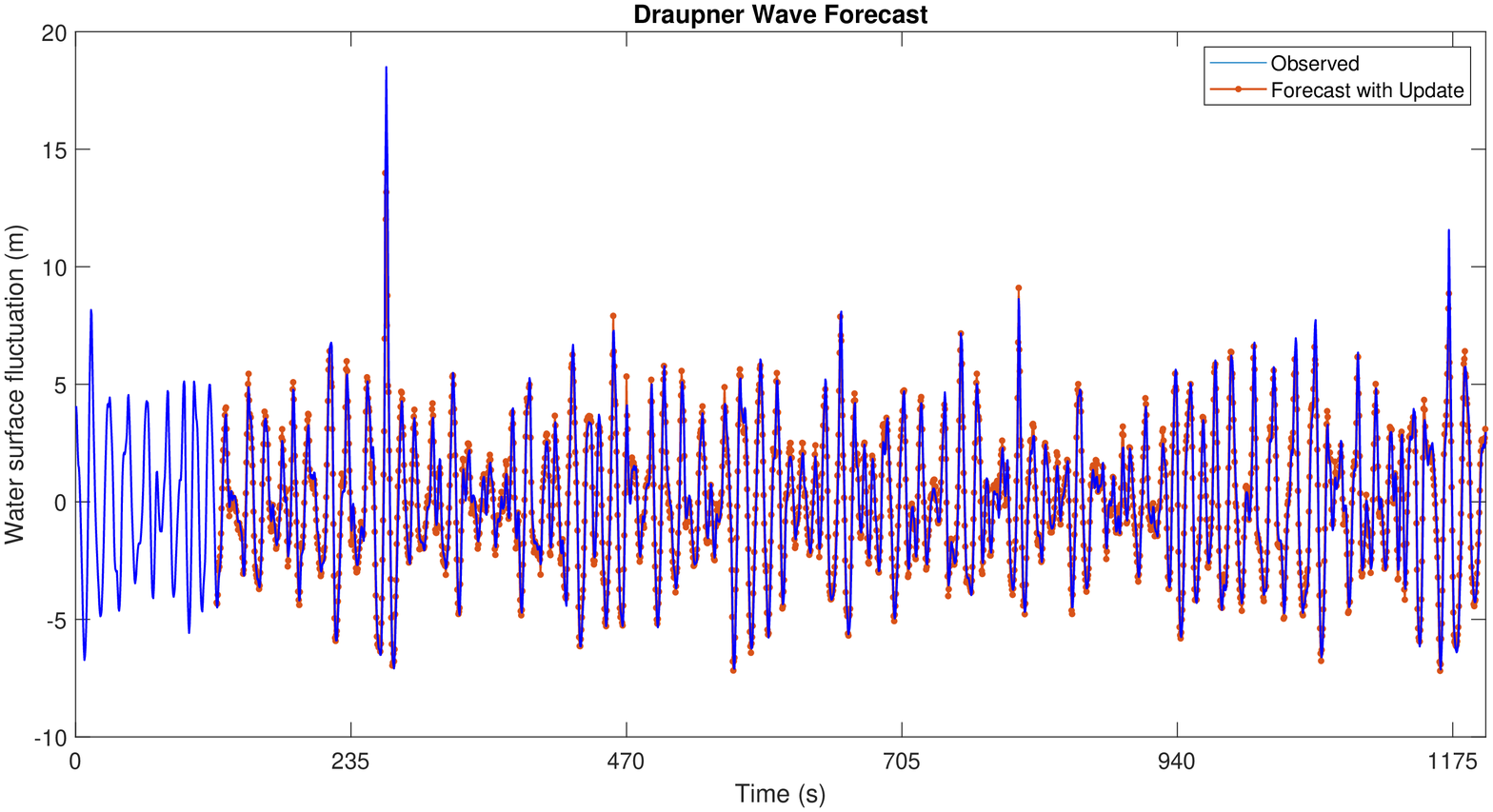}
  \end{center}
\caption{\small Comparison of the self-localized single soliton solutions of the KEE and NLSE}
  \label{fig9}
\end{figure}

\begin{figure}[htb!]
\begin{center}
\hspace*{-0.7cm}  
   \includegraphics[width=4.2in]{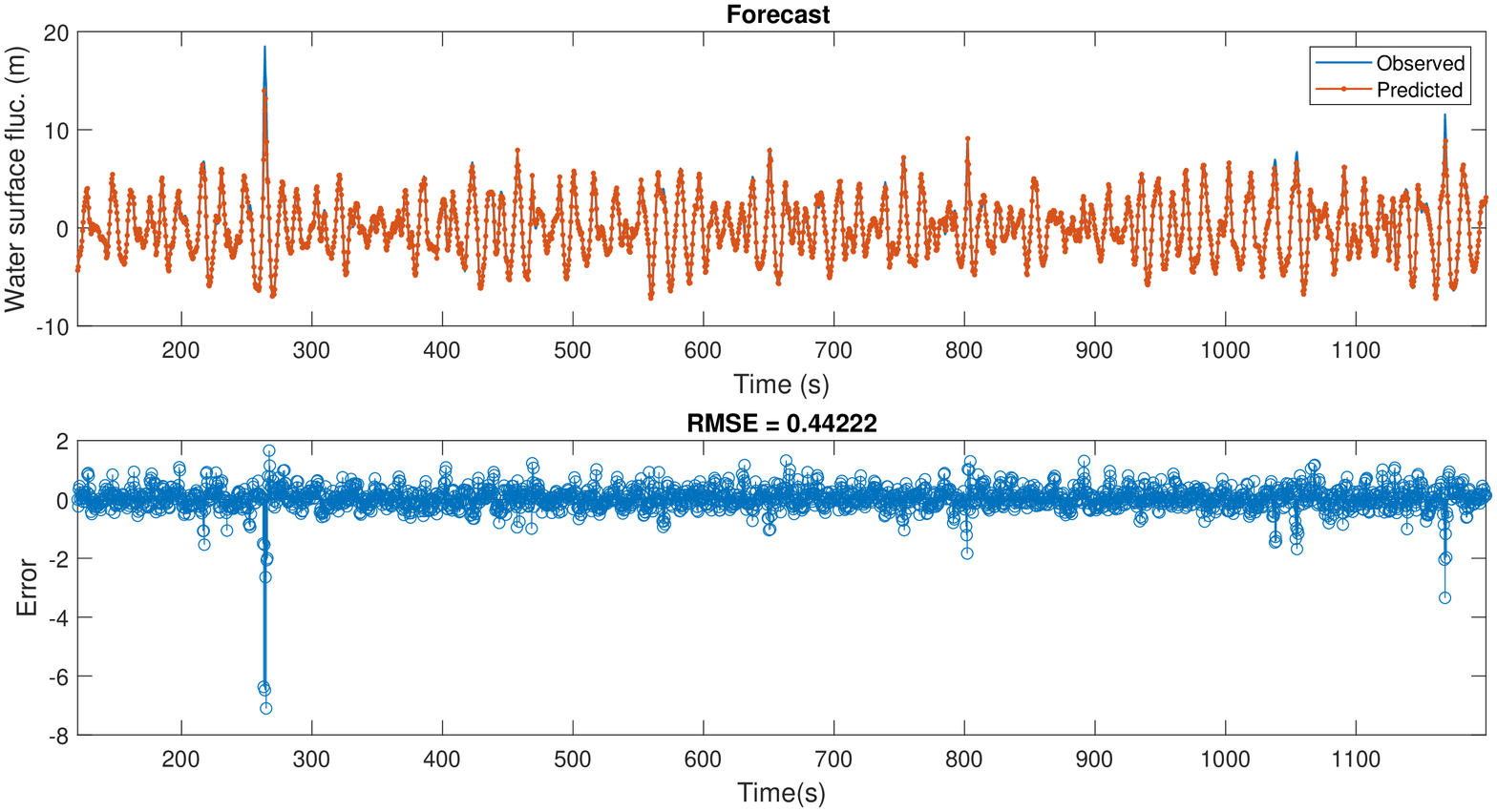}
  \end{center}
\caption{\small Comparison of the self-localized single soliton solutions of the KEE and NLSE}
  \label{fig10}
\end{figure}

In this section we turn our attention to the original Draupner rogue wave data, without any time reversing and analyze the original set as it is recorded. We use the first 10 \% of the data as our training set in LSTM network to test the remaining 90 \% of the water fluctuation time series. The results are depicted Fig.~\ref{fig9} and Fig.~\ref{fig10}.

As depicted Fig.~\ref{fig9} and Fig.~\ref{fig10}, LSTM can effectively be used to predict the Draupner rogue wave data using only the initial 10 \%  as the training set. The LSTM predicts a high wave at t=264s with a height around 21.5m and this wave can be classified as a rogue wave since it exceeds the $2H_s$ in the field, where $H_s$ is the significant wave height in the chaotic wavefield. The recording of the Draupner data indicates that at time t=264s of the time series, the dangerous peak had a waveheight of approximately H=26m. As in the previous cases explained above, the LSTM slightly underestimates the peak waveheight, however can be successfully used to predict the emergence time of catastrophic peak with an early warning time on the order of 264-124=140s, which is more than 2 minutes. With the improvement of the rogue wave sensing systems, existing data sets and deep learning algorithms; early warning time scales reported in this paper can be significantly improved in near future. Furthermore, it is also possible to improve the results presented in this paper using a combined deep learning and spectral analysis and/or compressive sensing approach, for example by using the results presented in \cite{Akhmediev2009a, Akhmediev2011, BayPRE1, BayPRE2, BayPLA, Bay_arxEarlyDetectCS}. LSTM network and other deep learning algorithms can also be used as time series forecasting and hindcasting tools to predict and complete missing rogue wave data in the fields such as hydrodynamics, optics, structural and machine vibrations, Bose-Einstein condensation, chaotic dynamics and gravity, just to name a few.

\section{\label{sec:level1}Conclusion and Future Work}

In this paper we have proposed and applied deep learning for the early detection of the Draupner rogue wave. More specifically, we have showed that LSTM based deep learning strategy can significantly improve the reported early warning times from order of seconds to order of minutes, at least. This is a major step forward which will certainly enhance the safety of the marine engineering including marine travel and offshore drilling. With the advance of the extreme event sensing systems and recorded data, the deep learning based methodology proposed in this paper is expected to produce even more successful results in predicting the emergence times and wave heights of rogue waves and other similar extreme event.  Additionally, existing spectral and nonlinear time series analysis tools can be used together with the deep learning algorithms leading to hybrid approaches to improve prediction time and predicted wave height.  Our results can also be used to predict various rogue wave and extreme phenomena in fields including but are not limited to hydrodynamics and marine engineering, optics, finance, Bose-Einstein condensation, mechanical vibrations and gravity just to name a few.

\section*{\label{sec:level1}Acknowledgment}
The author thanks Prof. Sverre Haver for providing access to the Draupner wave data.

\end{document}